\documentclass[journal]{IEEEtran}

\usepackage{cite}
\usepackage{graphicx}
\usepackage[cmex10]{amsmath}
\usepackage{amsfonts}
\usepackage{algorithmic}
\usepackage{multirow}
\usepackage{textcomp}
\usepackage{color}

\begin{document}

\title{Intelligent Wireless Communications Enabled by Cognitive Radio and Machine Learning}

\author{Xiangwei Zhou$^{1,*}$, Mingxuan Sun$^{1}$, Geoffrey Ye Li$^{2}$, and Biing-Hwang (Fred) Juang$^{2}$\\
$^{1}$School of EECS, Louisiana State University, Baton Rouge, LA 70803, USA\\
$^{2}$School of ECE, Georgia Institute of Technology, Atlanta, GA 30332, USA\\
$^{*}$Corresponding author, email: xwzhou@lsu.edu
\thanks{\textcopyright 2018 IEEE.}
}

\maketitle

\begin{abstract}
The ability to intelligently utilize resources to meet the need of growing diversity in services and user behavior marks the future of wireless communication systems. Intelligent wireless communications aims at enabling the system to perceive and assess the available resources, to autonomously learn to adapt to the perceived wireless environment, and to reconfigure its operating mode to maximize the utility of the available resources. The perception capability and reconfigurability are the essential features of cognitive radio while modern machine learning techniques project great potential in system adaptation. In this paper, we discuss the development of the cognitive radio technology and machine learning techniques and emphasize their roles in improving spectrum and energy utility of wireless communication systems. We describe the state-of-the-art of relevant techniques, covering spectrum sensing and access approaches and powerful machine learning algorithms that enable spectrum- and energy-efficient communications in dynamic wireless environments. We also present practical applications of these techniques and identify further research challenges in cognitive radio and machine learning as applied to the existing and future wireless communication systems.
\end{abstract}


\IEEEpeerreviewmaketitle

\section{Introduction}
Global mobile data traffic has grown significantly over the past years. In terms of monthly volume,
it grew from 400 petabytes/mo in 2011 to 7.2 exabytes/mo at the end of 2016, and is predicted to
reach 49 exabytes/mo by 2021 \cite{CISCO2017}. In addition, by 2020, the smartphone traffic will exceed the PC
traffic and mobile devices will account for two-thirds of the total IP traffic \cite{CISCO2016}. Along with the
remarkable growth in data traffic, new applications of wireless communications, such as wearable
devices and \emph{Internet of Things} (IoT) \cite{bandyopadhyay2011internet}, continue to emerge and generate even more data traffic.
With the exploding wireless traffic, applications, and device diversity, the future wireless
communication systems must embrace intelligent processing 
to address the universal scarcity in
spectrum and energy resources. This has led to research on cognitive radio \cite{Haykin05,Ma09} and machine learning \cite{alsheikh2014machine,jiang2017machine}, both of which form the pillars to support the intelligent processing requirements of wireless communication systems.

Intelligent processing in a wireless communication system encompasses at least
the following: 1) the perception capability, 2) reconfigurability, and 3) the learning capability.

The perception capability enables wireless environment awareness and is one of the most
important features in intelligent wireless communications. As a key component of cognitive
radio \cite{Haykin05,Ma09}, it allows the wireless operation of a device to adapt to its environment and potentially maximize
the utility of the available spectrum resources. The perception capability is afforded by spectrum
sensing \cite{cabric2004implementation,lu2012ten}, which in a narrow sense determines the spectrum availability. Many basic
spectrum sensing techniques have been proposed, including matched filter detection, energy
detection, feature detection, and covariance-based detection \cite{cabric2004implementation,zeng2009eigenvalue}. Advanced spectrum
sensing techniques to cope with various scenarios, such as cooperative spectrum sensing \cite{ganesan2007cooperative,Guru:ITW2,ma2008soft,zhou2010probability}, wideband spectrum sensing \cite{Tian}, and sequential spectrum sensing \cite{xin2014low}, have also been studied
over the last decade. In a broad sense, spectrum sensing can be regarded as a paradigm for
wireless environment perception. From this perspective, multi-dimensional spectrum sensing \cite{do2010joint}, spectrum measurements \cite{zhao2006radio}, and interference sensing \cite{zhang2016interference} have been considered in the recent literature. Since spectrum sensing requires resources at the sensing nodes, efficient scheduling of spectrum sensing \cite{Anh_Tuan_Hoang:Adaptive_Joint_Scheduling_of_Spectrum_Sensing10,zhou2009probability} has been discussed to balance the time, bandwidth, and power spent in between sensing and transmission.

To adapt to the surrounding environments, intelligent wireless devices need to be reconfigurable in addition to being able to perceive the environment. Reconfigurability is achieved by dynamic
spectrum access and optimization of operational parameters \cite{xing2007dynamic}. Based on the available information on the wireless environments and particular regulatory constraints, dynamic spectrum access techniques can be classified as interweave, underlay, overlay, and hybrid schemes \cite{srinivasa2007cognitive}. The main reconfigurable parameters of these schemes in the physical layer include waveform, modulation, time slot, frequency band, and power allocation. Given different levels of perception capability, various designs of spectrum access have been proposed \cite{peha2005approaches}. To achieve high performance, such as the throughput, and to satisfy certain constraints, such as the qualify-of-service requirements, different optimization algorithms \cite{zhang2014resource,zhou2010probabilistic}, including graph-based and market-based approaches, have been developed. The main challenges on the issue, including imperfect information, real-time requirements, and complexity limitations, have been considered. With the exploding number of wireless devices consuming a large amount of energy, energy efficiency also becomes important for dynamic spectrum access and resource optimization. Therefore, it has received increased attention recently \cite{xie2012energy}.

Resources in the wireless environments recognized by the perception capability
and reconfigurability design are characterized in a slew of factors, such as frequency band, access
method, power, interference level, and regulatory constraints, to name a few. Interactions among these
factors in terms of how they impact on the overall system utility are not always clearly known. As we try to maximize the utility of the available resources, the system complexity may thus be already daunting and can be further compounded
by the diverse user behaviors, thereby calling for a proper decision scheme that would help
realize the potential of utility enhancement. Modern machine learning techniques \cite{alsheikh2014machine,jiang2017machine,sun2012estimating,lee2012automatic} would find ample
opportunities in this particular application \cite{thilina2013machine,tsagkaris2008neural}. The learning capability enables
wireless devices to autonomously learn to optimally adapt to the wireless environments. In addition to the traditional machine learning approaches that use offline data, i.e., data collected in the past, to train models, efficient and scalable online learning algorithms that can train and update models continuously using real-time data are of great interest and have been successfully applied in various domains, including web search \cite{das2007google} and cognitive radio networks \cite{zhang2015spectrum}.

Machine learning algorithms are being developed at a fast pace. Both supervised and unsupervised learning algorithms have been used to address various
problems in wireless communications. Different from the standard supervised learning,
reinforcement learning has been found useful to maximize the long-term system performance and
to strike a balance between exploration and exploitation \cite{galindo2010distributed}. In addition, deep learning has emerged as a powerful approach to achieving superior and robust performance directly from a vast
amount of data, and therefore has great potential in wireless communications \cite{cui2015deep}.

\begin{figure} \centering
\includegraphics[width=0.35\textwidth]{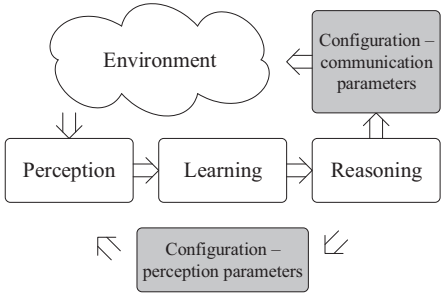}
\caption {Cognitive radio and machine learning in intelligent communications.} \label{overview1}
\end{figure}

Different from the scope of existing surveys in this area, we provide in this article a comprehensive overview of the development of cognitive radio 
and machine learning; in particular, we elaborate on their relationships and interactions in their roles towards achieving intelligent wireless communications as shown in Figure \ref{overview1}. Moreover, we consider spectrum and energy efficiency, both of which are important characteristics of intelligent wireless communications, rather than only focusing on improving spectrum efficiency as most other overview papers on cognitive radio do.

The rest of this paper is organized as follows. We describe the state-of-the-art of cognitive radio technology, covering spectrum
sensing and access approaches that perceive and adapt to the wireless environments in Sections \ref{cog} and \ref{rec}, respectively. In Section \ref{mac}, we present
powerful machine learning algorithms that enhance the perception capability and reconfigurability
in wireless communications. We discuss practical applications of these techniques to wireless communication systems, such as heterogeneous networks and \emph{device-to-device} (D2D) communications, and further elaborate some research challenges and likely improvements in
future intelligent wireless communications in Section \ref{app}. Finally, we conclude the paper in Section \ref{con}.

\section{Spectrum Sensing and Environment Perception} \label{cog}
The perception capability is the system's ability to detect and assess the parameters that exist in the wireless environment, ranging from the spectrum availability to the power consumption and reserve level during operation. It is one of the most
important features in intelligent wireless communications. As a key component of cognitive
radio \cite{Haykin05,Ma09}, it allows the wireless operation of a device to adapt to its environment and potentially maximize
the utility of the available spectrum resources.
In this section, we focus on the techniques associated with the perception capability. We start with an introduction to spectrum sensing, including its basics and techniques for determining the spectrum availability. We review different categories of spectrum sensing methods, such as local and cooperative spectrum sensing, narrowband and wideband spectrum sensing, block and sequential spectrum sensing, to cope with various scenarios in wireless communications. We then extend our discussion to environment perception, including multi-dimensional spectrum sensing, spectrum measurements and statistical modeling, and interference sensing and modeling, to enhance the intelligence in future wireless communication systems. We further include spectrum and energy efficiency considerations in wireless environment awareness, such as scheduling of spectrum sensing.

\subsection{Spectrum sensing}
The perception capability is mainly afforded by spectrum sensing \cite{cabric2004implementation,lu2012ten}, which in a narrow sense determines the spectrum availability at a particular time and geographical location.
For a particular frequency band, spectrum sensing decides between two hypotheses, ${\cal H}_0$ and ${\cal H}_1$, corresponding to the absence and the presence of the licensed user signal, respectively. 
With the spectrum sensing capability, an unlicensed cognitive radio user, also called a secondary user, can utilize the spectrum resources when a licensed user, also called a primary user, is absent or inactive.
The spectrum sensing performance is usually characterized by the probabilities of detection and false alarm. The probability of detection is
the probability that the decision is ${\cal H}_1$ while ${\cal H}_1$ is true; the probability of false alarm is
the probability that the decision is ${\cal H}_1$ while ${\cal H}_0$ is true. 
It is desirable to achieve a large probability of detection to enable efficient spectrum exploitation and a small probability of false alarm to limit or avoid undue interference to the licensed operation. In practice, spectrum sensing needs to strike a balance between the probabilities of detection and false alarm, as in typical hypothesis testing tasks where a proper operating point must be chosen.

\subsubsection{Basic approaches}
Many basic spectrum sensing approaches have been proposed, including matched filter detection, energy detection, feature detection, and covariance-based detection \cite{Ma09,cabric2004implementation,Poor,Ghozzi:Cyclostatilonarilty06,zeng2009eigenvalue}. The matched filter detector \cite{cabric2004implementation} correlates the received signal with a known copy of the licensed user signal to maximize the received \emph{signal-to-noise ratio} (SNR). Under \emph{additive white Gaussian noise} (AWGN), it is optimal given a known signal. However, it can only be applied when the patterns of the licensed user signal, such as preambles, pilots, and spreading sequences, are known to the secondary cognitive radio user. Energy detection \cite{Poor}, in contrast, simply compares the energy of the observed signal with a threshold and decides on the presence or the absence of the licensed user signal. It does not require a priori knowledge of the licensed user signal but is susceptible to the uncertainty of noise power level \cite{mariani2011effects}. Feature detection \cite{Ghozzi:Cyclostatilonarilty06} analyzes cyclic autocorrelation of the received signal. It can differentiate the licensed user signal from the interference and noise and even works in very low SNR regimes. Covariance-based detection \cite{zeng2009eigenvalue} utilizes the property that the received signal is usually correlated as a result of the dispersive channels, the use of multiple receive antennas, or oversampling, thus providing differentiation from the noise. It can be used without the knowledge of signal, noise power, and detailed channel properties.

\subsubsection{Local and cooperative spectrum sensing}
\begin{figure} \centering
\includegraphics[width=0.35\textwidth]{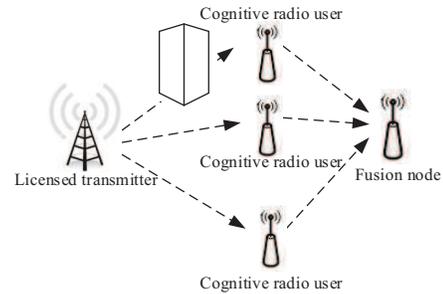}
\caption {Hidden node problem and cooperative spectrum sensing.} \label{ccs}
\end{figure}

Local spectrum sensing performed at a single cognitive radio user does not always render a satisfactory performance given noise uncertainty and channel fading \cite{tandra2008snr,molisch2009propagation}. For example, a cognitive radio user may not detect the licensed user signal shadowed by a high building as shown in Figure \ref{ccs}, which is known as the hidden node problem. 
If multiple cognitive radio users collaborate in spectrum sensing, the possibilities of detection error can be reduced with the introduced spatial diversity \cite{ghasemi2005collaborative}. It has been shown that cooperative spectrum sensing can also reduce the detection time needed at an individual cognitive radio user \cite{ganesan2007cooperative,Guru:ITW2}. In cooperative spectrum sensing, each cognitive radio user first independently performs local spectrum sensing and then sends its sensing data to a fusion node; according to its received information, the fusion node makes a decision on the presence or absence of the licensed user signal, as shown in Figure \ref{ccs}. A straightforward form is to send and combine the signal samples received by individual cognitive radio users \cite{Cardoso}. To reduce the required bandwidth, each user may instead send summary statistics, such as the locally observed energy or quantized sensing data \cite{ghasemi2005collaborative,ma2008soft,zhou2010probability}. In \cite{unnikrishnan2008cooperative}, the correlation among individual sensing results is further considered. The weight design for cooperative spectrum sensing under practical channel conditions and link failures is discussed in \cite{zhang2015distributed}.

\subsubsection{Narrowband and wideband spectrum sensing}
While many existing spectrum sensing methods focus on exploiting spectral opportunities over narrow frequency ranges, spectral opportunities over a wide frequency range are of great importance for wireless environment awareness and intelligent wireless communications. Different from narrowband spectrum sensing that makes a single decision for the entire frequency band of interest, wideband spectrum sensing identifies the availabilities of multiple frequency bands within the wideband spectrum. Wideband spectrum sensing can be categorized into Nyquist and sub-Nyquist approaches.
To reduce the implementation complexity, sub-Nyquist wideband sensing is introduced based on the compressive sensing technique. It allows the use of a sampling rate much lower than the Nyquist rate and thus fewer observations in comparison with its Nyquist counterpart \cite{tian2012cyclic}. 

\subsubsection{Block and sequential spectrum sensing}
Most spectrum sensing algorithms require a prescribed number of samples of the received signal for the hypothesis testing task, which is referred to as block spectrum sensing. In this case, a given time slot is provided for spectrum sensing. In some applications, the decision on the presence or absence of a licensed user signal needs to be made as quickly as possible using a variable number of samples, targeting a given probability of detection. Based on the sequential testing methodology introduced in \cite{wald1973sequential}, sequential spectrum sensing \cite{xin2014low} can be applied in such a case, where received signal samples are taken sequentially and the decision can be made as soon as the required detection reliability is satisfied. 
Sequential spectrum sensing is also employed for cooperative spectrum sensing \cite{yilmaz2012cooperative} and wideband spectrum sensing \cite{haghighi2010wideband}.

\subsection{Environment perception}
Spectrum sensing is an important component in wireless environment perception. Conventional spectrum sensing focuses on the spectral opportunities in frequency bands not being used at a particular time and geographical location. The broad sense of spectral opportunities nonetheless can be further expanded beyond the conventional concept of unused spectrum to other possibilities, such as shared spectrum as long as no harmful interference is introduced by the augmented spectral use. Therefore, multi-dimensional spectrum sensing that creates more spectral opportunities has become a subject of great interest lately. Furthermore, effective utilization of the spectrum resources is often enhanced by proper prediction of the spectral availability and thus it is advantageous and necessary to keep track of the past spectrum usage pattern over larger time and geographical scales. For this purpose, spectrum measurements and statistical modeling can be used \cite{zhao2006radio}. To fully explore the expanded paradigm of spectral opportunities, interference sensing \cite{zhang2016interference} has also been considered in the recent literature to address the interference factor that limits the potential spectrum reuse.

\subsubsection{Multi-dimensional spectrum sensing}
To allow wireless communication systems to operate in the same frequency band, it is desirable to avoid interference at the particular time and geographical location. Conventional spectrum sensing schemes intend to achieve this goal by identifying either temporal or spatial spectral opportunities. However, joint spatial-temporal opportunities can be exploited to further enhance spectrum efficiency. In \cite{do2010joint}, a joint spatial-temporal spectrum sensing scheme is proposed and the performance benefit over spatial-only or temporal-only spectrum sensing is analyzed. Furthermore, a geolocation database is used in \cite{wang2014spatial} together with spectrum sensing to better capture the joint spatial-temporal spectral opportunities.

Note that other information beyond spectrum occupancies, such as the SNRs, channel states, and modulation and coding schemes, can be acquired with parameter estimation algorithms to help exploit the spectral opportunities. 
The data-aided SNR estimation solutions and approximations are derived for high and low SNR cases over Rayleigh fading channels in \cite{abeida2010data}.
In \cite{rashad2007efficient}, new pilot patterns using subcarriers free from the interference are proposed for the \emph{orthogonal frequency-division multiplexing} (OFDM)-based cognitive radio and the channel estimation is studied based on the cascaded 1-dimensional Wiener filter. 
Automatic modulation classification exploiting the cyclostationarity property of the modulated signals is proposed in \cite{ramkumar2009automatic}. 
As shown in \cite{tsakmalis2014automatic}, the secondary user can change its transmitting power and limit the induced interference adaptively when it detects the modulation scheme of the primary user. 

\subsubsection{Spectrum measurements and statistical modeling}
A fundamental key to environment perception is the understanding of the historical and statistical properties of the spectrum occupancy.
The spectral opportunities in Chicago are studied in \cite{roberson2006spectral}, which demonstrates the potential use of cognitive radio technology for improved spectrum efficiency. In \cite{datla2009spectrum}, a framework for collecting and analyzing spectrum measurements is provided and evaluated. A statistical spectrum occupancy model in time and frequency domains is designed in \cite{ghosh2010framework}, where the first- and second-order parameters are determined from actual RF measurements. In \cite{wellens2010lessons}, a spectrum measurement setup is presented with lessons learned during the measurement activities. A model for the duty cycle distribution of spectrum occupancy is introduced and the impact of duty cycle correlation in the frequency band is discussed. The drawback of Poisson modeling of licensed user activities is considered in \cite{canberk2011primary}, in which a new model is introduced to account for the correlation in the licensed user statistics. A novel spatial modeling approach is proposed in \cite{lopez2017space} to address the problem of modeling the spectrum occupancy in the spatial domain. In addition, the radio environment map is investigated in \cite{zhao2006radio} to act as an integrated database consisting of multi-domain information.

\subsubsection{Interference sensing and modeling}
Interference temperature was proposed by the \emph{Federal Communications Commission} (FCC) as an indicator to guarantee minimal interference to the licensed users \cite{federal2003establishment}. While this concept is no longer popular nowadays, interference sensing and modeling are still an important aspect of environment perception for efficient and intelligent wireless communications.
In \cite{ghasemi2008interference}, the distribution of aggregated interference from cognitive radio users to a licensed user is characterized in terms of the sensitivity, transmitted power, and density of the cognitive radio users as well as the propagation environment. This statistical model can help design system-level parameters based on the interference constraint. The statistical behavior of the interference in cognitive radio networks is also studied in \cite{rabbachin2011cognitive} based on the theory of truncated stable distributions. The effect of power control is included in the discussion.

\subsection{Spectrum and energy efficiency considerations}
Since spectrum sensing requires resources at the sensing nodes, efficient scheduling of spectrum sensing is very important to balance the time, bandwidth, and power spent in between sensing and transmission. Note that periodic spectrum sensing is commonly used to avoid interference with licensed users that may suddenly appear in the middle of cognitive communications. On one hand, cognitive radio users may spend a lot of time on sensing activities rather than data transmission if spectrum sensing activities are scheduled quite often. On the other hand, available spectrum opportunities may not be quickly discovered if spectrum sensing activities are scheduled too sporadically. As a result, the spectrum efficiency relies not only on the spectrum sensing technique itself but also on how spectrum sensing activities are scheduled.

Typically, each frame consists of both a sensing block and an inter-sensing block \cite{Modeling}. The ratio of the block lengths represents the frequency that sensing activities are scheduled, and thus is a key design parameter for spectrum sensing scheduling. The optimization of spectrum sensing scheduling has been studied intensively for reliability-efficiency tradeoff \cite{DT,MACS,ST}. Sensing block length optimization is investigated in \cite{DT} and \cite{MACS} to enhance the spectrum efficiency of a cognitive radio user utilizing a single licensed channel and multiple licensed channels, respectively. The optimal sensing block length is determined to maximize the achievable throughput for the cognitive radio user under the constraint that the licensed users are sufficiently protected. Similarly, the optimal inter-sensing block length is considered in \cite{ST,zhou2009probability}. In \cite{Anh_Tuan_Hoang:Adaptive_Joint_Scheduling_of_Spectrum_Sensing10}, the optimization of both sensing and inter-sensing block lengths is studied. 

For better energy efficiency rather than spectrum efficiency, the optimization of inter-sensing block with data transmission is addressed in \cite{li2011energy}. To minimize the energy consumed in cooperative spectrum sensing, the sensor selection and optimal energy detection threshold are discussed in \cite{ebrahimzadeh2015sensor}. The throughput and energy efficiency tradeoff in cooperative spectrum sensing is further studied in \cite{ejaz2015energy}.

Moreover, the channel sensing order also affects the efficiency when there are multiple channels of interest. In \cite{jiang2009optimal}, a dynamic programming-based solution is provided for optimal channel sensing order with adaptive modulation, where both the independent and correlated channel occupancy models are considered.

\section{Reconfiguration: Spectrum Access and Resource Optimization} \label{rec}
To adapt to the surrounding environments, an intelligent wireless device needs to be reconfigurable in addition to being able to perceive the environment. Reconfigurability is achieved by dynamic
spectrum access and optimization of operational parameters \cite{xing2007dynamic}.

In this section, we focus on reconfigurability for intelligent wireless communications. We start with different types of dynamic spectrum access techniques, including interweave, underlay, overlay, and hybrid, as ways of coexistence in wireless networks with different levels of intelligence. Then we review resource optimization methods, including waveform and modulation design, resource allocation and power control, and graph- and market-based approaches. We will address uncertainties, imperfections, and errors, as well as other requirements and limitations. We further consider spectrum and energy efficiency tradeoff in intelligent reconfiguration, such as interference-aware spectrum access and resource optimization.

\subsection{Dynamic spectrum access techniques}
\begin{figure} \centering
\includegraphics[width=0.35\textwidth]{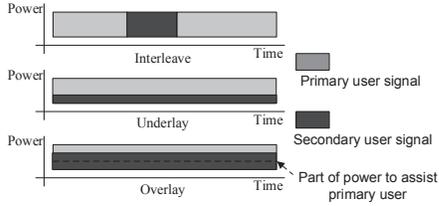}
\caption {Dynamic spectrum access with temporal spectrum sharing.} \label{dsa}
\end{figure}

Based on the available information on the wireless environments and particular regulatory constraints, dynamic spectrum access techniques can be classified as interweave, underlay, overlay, and hybrid schemes \cite{srinivasa2007cognitive}, as illustrated in Figure \ref{dsa}. As the original motivation for cognitive radio, secondary users exploit gaps in time, frequency, space, and/or other domains that are not occupied by primary users in the interweave paradigm. Ideally, interference is avoided in this paradigm since no user activities are found in spectrum holes. In practice, there may still be minor interference to the primary users with reliable spectrum sensing. In the underlay paradigm, secondary users are allowed to transmit simultaneously with primary users over the same frequency band if the interference generated by the secondary transmitters at the primary receivers is within some acceptable level that is usually very restrictive. In the overlay paradigm, secondary users are also allowed to transmit simultaneously with the primary users over the same frequency band. However, the interference generated by a secondary transmitter at a primary receiver in overlay communications can be offset by using part of the power of the secondary user to assist the transmission of the primary user. The hybrid paradigm \cite{sharma2014hybrid} combines some of the above paradigms to overcome their drawbacks.

\subsection{Resource optimization}
The main reconfigurable physical-layer parameters include waveform, modulation, time slot, frequency band, and power allocation. Given different levels of perception capability, various designs of spectrum access have been proposed \cite{peha2005approaches}. To achieve high performance, such as the throughput, and satisfy certain constraints, such as the qualify-of-service requirements, different optimization algorithms, including graph- and market-based approaches, have been developed \cite{zhang2014resource,zhou2010probabilistic}. The main challenges on the issue, including imperfect information, real-time requirements, and complexity limitations, have been considered in the recent literature.

\subsubsection{Waveform and modulation design}
To enhance the spectrum usage and minimize the interference to the primary users, the design of waveform and modulation for the secondary users can be optimized. In the underlay spectrum access, the secondary users can apply \emph{ultra wideband} (UWB) waveforms and optimize the pulse width and position \cite{zhang2006multiple}. In the overlay spectrum access, OFDM is an attractive transmission technique \cite{Weiss04} that can flexibly turn on or off tones to adapt to the radio environments. Meanwhile, with \emph{orthogonal frequency-division multiple access} (OFDMA) as the multiple access technique, non-adjacent sub-bands can be utilized with dynamic spectrum aggregation \cite{Chen08}. Due to the \emph{out-of-band} (OOB) leakage of the OFDM signal, spectrum shaping in either time or frequency domain becomes necessary to suppress the OOB radiation and thus to reduce the interference in the adjacent bands \cite{Weiss042}. To suppress the OOB radiation, a raised cosine window can be applied to the signal in the time domain \cite{Weiss042} but reduces system throughput because of the extension of symbol duration. Another time-domain method is adaptive symbol transition, which inserts extensions between OFDM symbols at the cost of system throughput reduction \cite{Mahmoud08}. In the frequency domain, the tone-nulling scheme \cite{Weiss04} simply deactivates OFDM subcarriers at the frequency band edges that introduce the most OOB emission to the adjacent bands. Furthermore, the active interference cancellation scheme \cite{Yamaguchi04} adaptively inserts cancelling tones at the frequency band edges to enable deep spectrum notches, which, however, is computational intensive at the transmitter. Similar schemes that suppress the OOB radiation based on the transmitted data include multiple-choice sequence \cite{Ahmed08}, selected mapping \cite{Ohta10}, subcarrier weighting \cite{Cosovic06}, and spectral precoding \cite{Chung08}. 

\subsubsection{Resource allocation and power control}
Resource allocation and power control have always been effective approaches for wireless networks. With the development of cognitive and intelligent wireless communication systems, various types of users may coexist in the same area and share the available spectrum resources through advanced dynamic spectrum access techniques. As a result, dynamic resource allocation and adaptive power control have been paid more and more attention recently. In the following, we discuss recent development of dynamic resource allocation and adaptive power control from different aspects, including information availability, allocation manners, requirements, and metrics.

\textit{Information availability:} The available information, such as \emph{channel state information} (CSI), is crucial in resource allocation and power control. For intelligent wireless communications, such information plays a more important role in dynamic resource allocation and adaptive power control. To explore the benchmark performance and facilitate the analysis, the availability of CSI is usually assumed. With the perfect CSI at the transmitter, the optimal power allocation strategies for cognitive radio users over fading channels is proposed and the corresponding ergodic capacity and outage capacity are analyzed in \cite{kang2009optimal}. With the assumption of perfect CSI, spectrum and energy efficient resource allocation for cognitive radio networks is considered in \cite{wang2013efficient} and \cite{wang2013energy}, respectively. 

In the dynamic wireless environments, obtaining the perfect information is not realistic, especially when a large number of parameters are taken into consideration for performance improvement in intelligent communications. In addition, precise information exchange also introduces unacceptable overhead. Therefore, more recent work considers dynamic resource allocation with partial CSI, imperfect spectrum sensing, and channel uncertainty. In \cite{le2008resource}, a resource allocation framework is proposed in cognitive radio networks with the use of the estimated CSI. In \cite{zhang2009robust}, a robust power allocation scheme is proposed to limit the interference to the primary user in cognitive radio networks with partial CSI. In \cite{zhou2010probabilistic}, resource allocation based on probabilistic information from spectrum sensing is derived for opportunistic spectrum access. With imperfect spectrum sensing and channel uncertainty, resource allocation in femtocell networks is addressed in \cite{zhang2016resource}, where the overall throughput of femtocell users is maximized under probabilistic constraints. In \cite{soltani2013chance}, chance-constrained uplink resource allocation is considered in downlink OFDMA cognitive radio networks with imperfect CSI. Moreover, the optimal resource allocation with average bit-error-rate constraint is proposed in \cite{wong2009optimal}.

\textit{Allocation manners:} With different architectures and scales of wireless networks, resource allocation may be in a centralized or distributed manner. In the centralized manner, a central controller has sufficient information to render globally optimal allocation and hence to achieve good performance. In \cite{gulbahar2012information}, a centralized spectrum and power allocation scheme achieves maximum information capacity in a multi-hop cognitive radio network via correlations of sensor data and energy adaptive mechanisms. To reduce spectrum sensing overhead and improve the spectrum efficiency, centralized dynamic resource allocation for cooperative cognitive radio networks is proposed in \cite{kumar2015dynamic}. However, resource allocation in the centralized manner faces some practical issues, including huge overhead for information exchanging, signal transmission delay, high computational complexity, and the scalability of the proposed algorithms. 

In distributed resource allocation, the aforementioned issues can be effectively alleviated. As a result, distributed resource allocation becomes the subject of recent research endeavor. In \cite{ngo2011distributed}, joint subcarrier assignment and power allocation optimizes the performance of an OFDMA ad hoc cognitive radio network distributively. The proposed distributed algorithms are with affordable computational complexity and reasonable performance. In \cite{ghorbel2016distributed}, a two-stage heuristic resource allocation scheme through a learning-based algorithm is designed. The dynamic spectrum allocation and adaptive power control are accomplished with the help of individual user observations in two separated stages. To balance the performance and practical issues, a four-phase partially distributed downlink resource allocation scheme is developed for a large-scale small-cell network in \cite{sadr2014partially}.

\textit{Requirements:} To satisfy various demands and application requirements, optimal resource allocation and power control can be designed in different ways. Fairness and outage probability of joint rate and power allocation for cognitive radio networks are studied in a dynamic spectrum access environment in \cite{kim2008joint}. Furthermore, resource allocation schemes with max-min and proportional fairness are proposed for cognitive radio networks in \cite{le2008resource}. With the proposed algorithms, the optimal solutions to the admission control problem for the primary users and the joint rate and power allocation for the secondary users can be obtained. To better manage the interference, a three-loop power control architecture is presented in \cite{yun2011adaptive}. Based on the feedback information, the proposed architecture determines the optimal maximum transmit power, the target \emph{signal-to-interference-plus-noise ratio} (SINR), and the instantaneous transmit powers of femtocell users. In \cite{chandrasekhar2009power}, a link adaption-based power control scheme is derived for two-tier femtocell networks. The optimal power allocation is obtained through solving the formulated reward-penalty link SINR problem. Meanwhile, a cellular link protection algorithm is proposed to alleviate the cross-tier interference to the cellular users. To accommodate the \emph{quality-of-service} (QoS) requirement, QoS provisioning spectrum resource allocation is proposed for cognitive heterogeneous networks and cooperative cognitive radio networks in \cite{alshamrani2011qos} and \cite{doost2014spectrum}, respectively. Moreover, delay-aware resource allocation is developed based on a Lagrangian dual problem in \cite{el2014joint}. With the fast development of intelligent wireless communications, dynamic resource allocation problems with different requirements need to be further explored.

\textit{Metrics:} With the explosive growth of wireless communications, the spectrum scarcity and energy consumption have been paid more and more attention. The most recent study on resource allocation and power control has been focusing on spectrum efficiency and energy efficiency metrics. In the IoT, thousands of devices and sensors are connected to the Internet wirelessly, resulting in more and more scarce spectrum resources. Therefore, the study on resource allocation for high spectrum efficiency, especially in dynamic spectrum sharing scenario, has drawn a lot of attention. In \cite{asghari2011resource}, an adaptive time and power allocation policy over cognitive broadcast channels is studied. A sensing-based optimal resource allocation scheme and a low-complexity suboptimal solution are proposed to maximize the spectrum efficiency. From the throughput perspective, a three-dimensional resource allocation optimization problem is solved via the proposed heuristic algorithms in \cite{bhardwaj2016enhanced}. The tradeoffs between performance and computational complexity of the proposed learning and optimization algorithms in dynamic spectrum access networks are analyzed. Due to the increasing energy consumption in wireless applications and services, the concept of green communications has been emphasized recently. Therefore, energy efficiency, as an important metric, has been extensively explored in resource allocation and power control. More details will be provided in the following subsection.

\subsubsection{Graph-based and market-based approaches}
Graph theory is a useful tool to model pairwise relationships between nodes. The most common application of graph theory to resource optimization is conflict graph, or interference graph, which describes the co-channel interference using nodes and edges. With the help of independent sets, groups of users allowed to use the same channel simultaneously without unacceptable interference can be identified. This feature benefits spatial spectrum reuse that significantly enhances spectrum efficiency. In \cite{Cohen2017Distributed}, a spatial channel selection game is proposed to increase spectrum efficiency with the use of conflict graph. In \cite{Kasbekar2012Spectrum}, spatial spectrum reuse is modeled as a price competition game among primary users, leading to a unique symmetric \emph{Nash Equilibrium} (NE) if the conflict graph of secondary users admits specific topologies defined as mean valid. In \cite{Wang2013Graph}, a peer-to-peer content sharing approach is proposed in vehicular ad hoc networks, in which a coalitional graph game is introduced to model the cooperation among vehicles and a dynamic algorithm is developed to find the best response network graph. In \cite{Xu2013Opportunistic}, a graphical game that describes channel selections for opportunistic spectrum access is proved to be a potential game and an NE, which minimizes the \emph{media access control} (MAC) layer interference. In addition, two uncoupled learning algorithms are proposed to approach the NE.

Market-based approaches of resource optimization treat spectrum resources as tradable items. These approaches give primary and secondary users motivations, usually opportunities of maximizing their own utilities, to participate in a predesigned spectrum sharing mechanism. The measure of utility varies in different scenarios. Common measures of utility are channel capacity and price of unit spectrum resource. The design of a spectrum sharing mechanism expresses the will of authority and the relationships among users are often involved in a game. Spectrum efficiency maximization is a common goal of a spectrum sharing mechanism using market-based approaches. In \cite{Gao2011MAP}, an auction process is introduced to implement dynamic spectrum access for secondary users when there are multiple channel holders. Assuming the existence of price competition among auctioneers systematically, the proposed multi-auctioneer progressive auction maximizes the spectrum efficiency. In \cite{Wang2014Truthful}, a truthful spectrum auction mechanism is proposed to allocate spectral resources according to both the demands and spectrum utilization. In \cite{Duan2014Cooperative}, dynamic spectrum access of secondary users is implemented as cooperative spectrum sharing under incomplete information. In the cooperative game, the secondary users act as relays of the primary user to exchange spectrum access time. By applying the contract theory, two optimal contract designs are proposed for weakly and strongly incomplete information scenarios. In \cite{Wu2014Revenue}, a two-layer game is proposed between a \emph{primary network operator} (PNO) and a \emph{secondary network operator} (SNO). In the top layer, the revenue sharing game is modeled as a Nash bargaining game, and both the PNO and SNO are benefited if they choose to cooperate. In the bottom layer, the resource allocation game is modeled as a Stackelberg game to determine the optimal spectrum allocation. These two layers improve iteratively and an equilibrium state exists. In \cite{Qian2011Spectrum}, an agent-based spectrum trading game considers the flexible demand of secondary users. In the case of a single agent, it is proved that there exists an optimal solution. In the case of multiple agents, the equilibrium of strategies of the agents can be obtained.

In some cases, the utility maximization of users does not result in spectrum efficiency maximization. In \cite{Zhang2016Location}, an evolutionary game is applied to modeling the pricing competitions among the primary users when the demands of the secondary users are related to channel prices. An evolutionary stable strategy is proved to exist when the primary users sell all their channels. However, the primary users have the opportunities to increase their payoffs by selling a portion of their channels. In \cite{Niyato2009Dynamics}, an adaptive spectrum sharing market between multiple primary and secondary users is introduced, in which the primary users adjust their prices and spectrum supplies and the secondary users change their channel valuations accordingly. By modeling the behavior of the secondary users as an evolutionary game and the competition of the primary users as a non-cooperative game, optimal strategies of both types of users are provided accordingly.

\subsection{Spectrum- and energy-efficient designs}
With the exploding number of wireless devices consuming a large amount of energy, energy efficiency is also important for dynamic spectrum access and resource optimization. Therefore, it has received increased attention recently, especially for battery-powered mobile devices. In \cite{hasan2009energy}, reliable power and subcarrier allocation in OFDM-based cognitive radio networks is studied from the energy efficiency perspective, where an energy-aware convex optimization problem is formulated and the corresponding optimal solution is obtained through a risk-return model. In \cite{devarajan2012energy}, user selection and power allocation schemes are proposed to reduce the energy consumption for a multi-user multi-relay cooperative communication system. A weighted power summation optimization for base and relay stations is formulated and solved. Furthermore, a multi-objective scheme that jointly considers the throughput performance and energy consumption is proposed to strike a balance between spectrum efficiency and energy efficiency.

Spectrum- and energy-efficient resource allocation schemes have been studied and proposed in various scenarios. However, simultaneously optimizing spectrum and energy efficiency is not possible in most cases \cite{li2011energy,hong2015optimal}. Therefore, the tradeoff between spectrum and energy efficiency plays an important role in resource allocation with different network architectures and requirements. For example, the increasing transmit power always improves spectrum efficiency but may reduce the energy efficiency in an interference-free environment. In an interference-limited environment, however, the increasing transmit power may decrease spectrum and energy efficiency at the same time. Moreover, the tradeoffs between spectrum and energy efficiency in downlink and uplink are not equivalent. The subcarrier allocation, power allocation, and rate adaption need to be jointly considered in the downlink while it is hard to perform joint optimization in the uplink. In addition to the energy consumption for data transmission, the energy consumption for spectrum sensing, information exchange, and the training of learning algorithms needs to be take into account and the corresponding tradeoff between spectrum and energy efficiency needs to be reconsidered.

Market-based spectrum sharing approaches can also help improve energy efficiency when the measure of utilities of secondary users is related to the transmit power. In \cite{Razaviyayn2011Stackelberg}, a decentralized Stackelberg pricing game is developed to find the optimal power allocation in the scenario of spatial spectrum reuse, such that utilities of the primary and secondary users are maximized. Two methods are proposed to solve the Stackelberg game in two different cases, i.e., active and inactive power constraints. In \cite{Huang2006Auction}, two auction mechanisms corresponding to two different pricing schemes are proposed for spatial spectrum reuse. When prices are set according to the received SINR, the auction mechanism leads to a weighted max-min fair allocation in terms of SINR. When prices are set according to the transmit power of users, the auction mechanism maximizes the total channel utility.

\section{Machine Learning for Intelligent Wireless Communications} \label{mac}
In this section, we focus on machine learning for intelligent wireless communications. Resources in the wireless environments recognized by the perception capability and reconfigurability design are characterized in a slew of factors, such as frequency band, access
method, power, interference level, and regulatory constraints, to name a few. Interactions among these factors in terms of how they impact on the overall system utility are not always clearly known. As we try to maximize the utility of the available resources, the system complexity may thus be already daunting and can be further compounded by the diverse user behaviors, thereby calling for a proper decision scheme that would help realize the potential of utility enhancement. Modern machine learning techniques \cite{alsheikh2014machine,jiang2017machine,sun2012estimating,lee2012automatic} would find ample
opportunities in this particular application \cite{thilina2013machine,tsagkaris2008neural}. Machine learning aims at providing a mechanism to guide the system reconfiguration, given the environment perception results and the device reconfigurability, to maximize the utility of the available resources. In other words, the learning capability enables
wireless devices to autonomously learn to optimally adapt to the wireless environments. 

Basic machine learning algorithms can be categorized into supervised and unsupervised learning. Reinforcement learning is emerging as a new category. Under each category, we will introduce specific learning models and discuss their applications in achieving intelligent communications. We further introduce the most recent development in machine learning, such as neural networks and deep learning, and discuss their potential for enhancing intelligent communications.

In our discussion, we consider different subcategories of machine learning algorithms based on their functionalities, such as support vector machines, Bayesian learning, k-means clustering, principal component analysis, and Q-learning. We will review the roles of different learning techniques in enhancing perception capability and reconfigurability, in which we specify the inputs, i.e., what to use, and the outputs, i.e., what to learn. For example, the required detection probability, observable wireless environment information, and available time or energy resource can be used to learn the selection of spectrum sensing methods and parameters while the available frequency bands, transmit power limit, and interference level can be used to learn the choice of channel assignment and power allocation. We evaluate and compare the strengths and limitations of different machine learning algorithms, to enable the choice of learning techniques and address various accuracy, complexity, and efficiency requirements of individual and cooperative tasks in centralized and decentralized wireless networks.

\subsection{Broad categories of machine learning algorithms}
Machine learning algorithms \cite{bishop2006pattern} learn to accomplish a task $T$ based on a particular experience $E$, the goal of which is to improve the performance of the task measured by a specific performance metric $P$ by exploiting the experience $E$. Depending on how to specify $T$, $E$, and $P$, machine learning algorithms are typically divided into three broad categories. Supervised learning accomplishes tasks by learning from examples provided by some external supervisor. Each training example consists of a pair of an input and an expected output/label, and the goal is to learn a function that predicts correctly the output for any input. 

In contrast to supervised learning, unsupervised learning algorithms generally assume that there are no labeled examples and the goal is to discover the hidden structure in the input. Typical algorithms include clustering algorithms that group inputs into a set of clusters and dimension reduction algorithms that reduce the dimensionality of the inputs. 

Furthermore, reinforcement learning emerges as a popular category, where an agent learns to perform a certain task, such as driving a car with minimal collisions, by interacting with a dynamic environment. In contrast to supervised learning, the agent obtains the feedback in terms of rewards only by interacting with the environment and learns on its own, which makes the reinforcement learning paradigm very useful for cases of decision making under uncertainty.

\subsection{Supervised learning}
Assume that there are $n$ training examples $\{(x_{1}, y_{1}),\ldots, (x_{n}, y_{n})\}$ available, where $x_{i} \in \mathcal{X}$ is an input of dimension $d$ and $y_{i} \in \mathcal{Y}$ is the corresponding output label. The goal of supervised learning is to find some function $f: \mathcal{X} \rightarrow \mathcal{Y}$ from a set of possible functions, such that $f$ not only approximates the relationship between input $\mathcal{X}$ and output $\mathcal{Y}$ encoded by the training examples but also generalizes well on unseen data. The learning task can be further divided into a regression task if the output is continuous or a classification task if the output is discrete. Taking classification as an example, an input $x_i$ can consist of the measurements from energy detection mentioned in Section \ref{cog} with $d$-dimension attributes, and the output is a binary variable indicating spectrum availability.

\begin{figure} \centering
\includegraphics[width=0.18\textwidth]{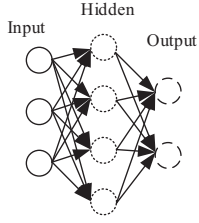}
\caption {Illustration of ANN. Each circular node represents a neuron and an arrow connects the output of one neuron to the input of another. Signals travel from the input layer to the output layer by traversing multiple hidden layers.} \label{ann}
\end{figure}

Popular supervised models include \emph{k-nearest neighbors} ($k$-NN or KNN), \emph{support vector machine} (SVM), \emph{Probabilistic graphical models} (such as Bayesian networks), and \emph{artificial neural network} (ANN). In particular, probabilistic graphical models such as \emph{hidden Markov model} (HMM) is designed to model the probability distributions of sequences of observations, e.g., measurements of time series. An ANN model, as shown in Figure \ref{ann}, can model any function regardless of its linearity between inputs and outputs given sufficient hidden layers and nodes in the network. Traditional ANNs usually construct networks with fewer than $5$ layers and the performance sometimes is not so good as other simpler models. As the accelerated \emph{graphics processing unit} (GPU) computing becomes more and more popular, larger volumes of training data become available, and more and more effective training algorithms are developed, deep learning~\cite{Lecun2015nature}, with constructions of deeper layers of a neural network, has been enjoying a major resurgence very recently, which we will discuss in detail later.

Since each specific model encodes different assumptions about the learning problem, they are different in terms of accuracy, computational efficiency, and types of applications. SVM with a linear kernel and logistic regression are both well-behaved classification algorithms and easy to train as long as the inputs are roughly linearly separable. SVM with non-linear kernel performs better for problems where the inputs might not be linearly separable. However, the models can be painfully inefficient to train, especially when the number of training examples, $n$, becomes increasingly large. Moreover, the neural network models are likely to perform well for most cases but are slower and harder to train. In contrast to all of the previous ``eager" learners, ``lazy" learners, such as KNN, do not learn a discriminative function from the training examples but simply ``memorize" all of them. However, the simplicity of KNN comes with a huge computational cost in prediction given new input, which involves searching for its nearest neighbors in the whole training set.

KNN and SVM can be used in intelligent wireless communications for cooperative spectrum sensing \cite{thilina2013machine}, where the input is the energy level estimation from a set of cognitive radio users and the output is the label for one of the two classes/hypotheses, corresponding to the absence and presence of the licensed user signal, respectively. The proposed cooperative spectrum sensing techniques based on KNN and SVM are shown to be more adaptive to the changing environments than the traditional methods. In addition, a computational efficient architecture is proposed for cooperative spectrum sensing using SVM, where the training process and online prediction can be operated independently. Specifically, the training process, regardless of its computational cost, can be performed in the background and the SVM model will only be updated when the radio environment changes. Whenever the cognitive radio network needs to identify the channel availability, energy features will be collected as input to the model and the model will generate output about the prediction of channel availability without too much delay.

SVM is also useful to perform classifications \cite{yang2010mac,hu2014mac,petrova2010multi} in wireless networks.
In \cite{yang2010mac}, SVM is used to classify contention-based or control-based MAC given the input features, i.e, the mean and variance of the received power at a cognitive radio terminal. From \cite{yang2010mac}, the model deployed by the cognitive radio terminals can classify \emph{time-division multiple access} (TDMA) and slotted ALOHA MAC protocols effectively. As the features related to the instantaneous received power are more distinctive between the two protocols when the new packet generating/arriving probability of the primary network increases, the classification accuracy improves. Similarly, SVM with different types of kernels is used in \cite{hu2014mac} to identify one of the four types of MAC protocols, including TDMA, \emph{carrier sense multiple access with collision detection} (CSMA/CA), pure ALOHA, and slotted ALOHA, according to the power features.
Multi-class SVM models are proposed in \cite{petrova2010multi} to classify seven distinct modulation schemes based on their spectral and statistical features, where the spectral features include the maximum of the spectral power density of the normalized centered instantaneous amplitude and the standard deviation of the absolute non-linear centered instantaneous phase, and the statistical features include higher-order statistics of the real part of the complex envelope.

A hierarchical SVM model is proposed in \cite{feng2012determination} to identify wireless network parameters, including the physical location in an indoor wireless network and channel noise level in a \emph{multiple-input multiple-output} (MIMO) wireless network. When there are a large number of transmit and receive antennas, the parameter identification may lead to search problems in a high-dimensional space. The proposed SVM based model is able to determine these parameters according to simple network information, such as the hop counts.
Five different models, including KNN and SVN, are used to predict a mobile user's specific usage pattern of data and location services~\cite{donohoo2014context}, which can further help optimize energy consumptions of mobile devices. Specifically, the input consists of spatial temporal context and device features, such as time, location, battery, and the number of running processes. The output of the classification is the on/off status of data and location services. Among the five strategies, the SVM achieves higher prediction accuracy and more energy savings with a minimal number of active users.

In \cite{jiang2014multi}, the multi-channel sensing and access problem is modeled as an Indian Buffet game, where the secondary users are customers and the primary channels are represented as a number of dishes in the restaurant. The proposed algorithm finds the perfect Nash equilibria of the subgames for the secondary users. To address the multi-channel sensing problem, a cooperative approach is used to estimate the channel state using Bayesian learning.
To maximize the spectrum utilization in cognitive radio networks \cite{zheng2013spectrum}, a Bayesian detector for multi-phase shift keying modulated primary user signals is proposed based on Bayesian decision rule. From \cite{zheng2013spectrum}, the Bayesian detector performs better than the traditional energy detector in terms of both spectrum utilization and secondary user throughput, especially in the high SNR regime.

The Bayesian learning techniques can also address the pilot contamination problem in massive MIMO systems \cite{wen2015channel}. The proposed approach outperforms the conventional estimators in both channel estimation accuracy and achievable rates when pilot contamination occurs.

An HMM model is constructed in \cite{choi2013estimation} to estimate the sojourn times of a primary user in both active and inactive states as well as the primary user signal strength, based on the sequence of the spectrum sensing results. The HMM is based on a two-state hidden Markov process, where the two states are whether the primary user signal is absent or not at each observation time in successive time frames. The parameters are estimated by extending the standard \emph{expectation-maximization} (EM) algorithm. The proposed algorithm can estimate the channel parameters well under certain conditions.

Variations of ANN models are used for spectrum sensing \cite{tang2010artificial,tumuluru2010neural,taj2011cognitive}. An ANN model is proposed in \cite{tang2010artificial} to predict binary channel status according to the features extracted from energy detection and cyclostationary feature detection. The proposed method can detect the signals well even if the SNR is low. By representing the channel status at every time slot as a time series \cite{tumuluru2010neural}, a \emph{multilayer feedforward neural networks} (MFNN) model is used to predict if the channel is busy or idle in the next slot based on the states in the previous $n$ slots. Compared with the HMM based approaches \cite{choi2013estimation}, the proposed approach can model the correlation between the current status and a large number of past observations more effectively. Instead of directly modeling the channel status, the model in \cite{taj2011cognitive} constructs a multivariate time series and predicts the evolution of RF time series data by exploring the cyclostationary signal features at every time slot and using a \emph{recursive neural network} (RNN).

There are also some applications of the ANN in the context of signal classification. An ANN model is used in \cite{popoola2011novel} to classify the signal into one of $12$ classes of analog and digital modulation schemes. The ANN models are also used for performance prediction and resource optimization. For example, the model in \cite{baldo2008learning,baldo2009neural} predicts the application-layer throughput of the mobile user, based on both environmental measurements, such as the packet rate, idle time, day of the week, and hour of the day, and parameter settings, such as the routing protocol being used via an MFNN. The proposed method for dynamic channel selection is implemented in IEEE 802.11 wireless networks, which demonstrates that the model can effectively predict the network performance with respect to the changes in the environment and dynamically select the best channels.

\subsection{Unsupervised learning}
\begin{figure} \centering
\includegraphics[width=0.18\textwidth]{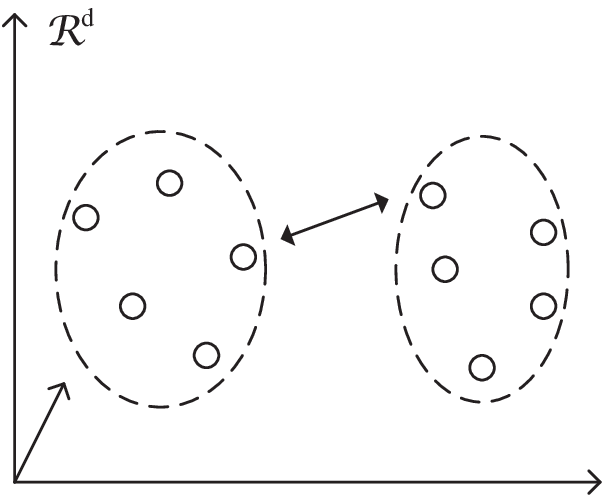}
\includegraphics[width=0.36\textwidth]{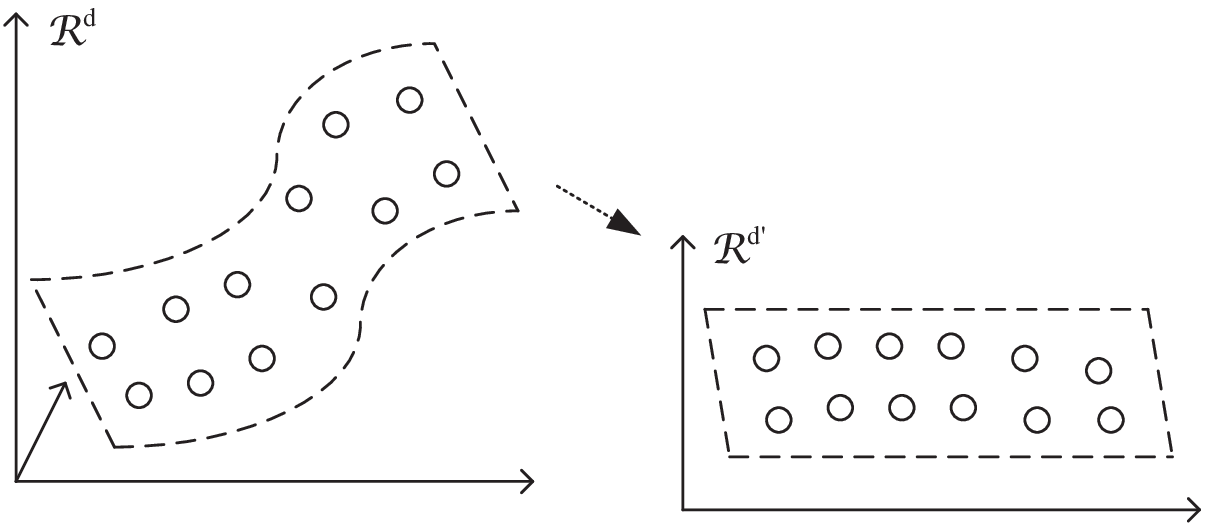}
\caption {Illustration of clustering (left) and dimensionality reduction (right).} \label{dr}
\end{figure}

As illustrated in Figure \ref{dr}, clustering, a typical unsupervised learning approach, groups $n$ observations $\{x_{1},\ldots, x_{n}\}$ to $k$ clusters, where $x_{i} \in \mathcal{X}$ is an observation of dimension $d$, so that the observations in the same cluster are more similar to each other and those in different clusters are less similar. Clustering has various applications in intelligent wireless communications. For example, small cells in heterogeneous networks can be clustered to avoid interference, the mobile users can be clustered to satisfy an optimal offloading policy, and the devices can be clustered to achieve high energy efficiency.

As one of the most popular clustering algorithms, centroid-based clustering, such as k-means,
assumes that there are $k$ clusters and each is associated with a centroid that is the average of all observations in the cluster. The goal is to find the clusters such that the sum of distances of the observations in the clusters to the centroids is minimized. 
In contrast to the k-means algorithm that assumes a fixed number of clusters, a \emph{Dirichlet Process Mixture Model} (DPMM) is a non-parametric Bayesian approach applied to clustering without predefining the number of clusters. Note that ``non-parametric" here implies a model whose parameters may change with observations, rather than a parameter-less model. The flexibility offered by non-parametric approaches, such as the DPMM, leads to a wide range of clustering applications that take account of the dynamic RF environments.

A cluster-based approach, ``HEED", is proposed in~\cite{younis2004heed} to group ad hoc sensor networks when channel allocation is fixed. As the radio spectrum usage paradigm tends to be open, the topology management algorithm in \cite{chen2007topology} solves the network formation problem in the cognitive radio context. The algorithm optimizes the cluster configurations to adapt to network and radio environment changes. To efficiently aggregate the source information under energy constraints, a distributed spectrum-aware clustering technique is developed in \cite{zhang2011distributed} to identify energy-efficient clusters and restrict the interference to the primary users in cognitive radio sensor networks. The goal is to identify a set of clusters to minimize the communication power, which is proved to be equivalent to minimizing the sum of squared distances between each node and its cluster center. A group-wise constrained clustering approach similar to k-means is proposed to minimize the intra-cluster distances with an additional spectrum-aware constraint. The proposed clustering technique is proved to be scalable and stable due to its quick convergence under dynamic primary user activities.

The DPMM can identify different types of wireless communication systems coexisting in the same frequency band \cite{shetty2009identifying}, where the number of systems may change over time. The observation here consists of features extracted from the received signal, including the center frequency and frequency spread after sensing. Since systems transmitting at different carrier frequencies will result in different observations, the DPMM only needs to group the observed data into different clusters, each representing a primary system that exists in a certain frequency band at a certain time. Similarly, the DPMM can be used to infer different types of signals, such as WiFi and Bluetooth signals, given their spectral and cyclic properties \cite{bkassiny2013multidimensional}.

Dimensionality reduction, as shown in Figure \ref{dr}, aims to transform observations of high-dimensional variables into meaningful low-dimensional representations without losing too much information. Specifically, dimensionality reduction techniques map high-dimensional observations $\{x_{1},\ldots, x_{n}\}$ into new low-dimensional representations $\{z_{1},\ldots, z_{n}\}$, where $x_{i} \in \mathcal{X}$ of dimension $d$, $z_{i} \in \mathcal{Z}$ of dimension ${d'}$, and $d' \leq d$. Classic methods include \emph{Principle component analysis} (PCA) and \emph{Independent component analysis} (ICA).
In contrast to PCA that projects observations to components that maximize the variance, the components of ICA have maximum statistical independence. Dimensionality reduction, including PCA and ICA, can be applied to various areas, such as signal denoising and separation.

Robust PCA is applied to recovering the low-rank covariance matrix of a signal corrupted by noise in cognitive radio networks \cite{hou2012spectrum}. In particular, assume that the received signal includes the desired signal and noise. As the covariance matrix of the white noise is diagonal and that of the signal is usually low-rank, the low-rank matrix can be extracted from the noisy sample covariance matrix with robust PCA. Therefore, the primary user signal can be found present if the difference between the recovered low-rank matrix and the original one is smaller than a predefined threshold, which is verified with both the simulated and captured digital television signals. The robust PCA can also be regarded as a denoising process for the sample covariance matrix in a similar way \cite{qiu2011cognitive}.
In \cite{qiu2011cognitive}, ICA is used in a smart grid scenario to separate wireless signals of smart utility meters from independent sources. The proposed model can be used to avoid channel estimation in each time frame and thus enhance the transmission efficiency. In addition, data security is preserved by avoiding wideband interference and eliminating jamming signals. Another example of ICA \cite{nguyen2013binary} is to decompose the observations of secondary users to mixtures of hidden binary primary user sources in cognitive radio networks. From \cite{nguyen2013binary}, the activities of up to $2m-1$ distinct primary users can be inferred given $m$ secondary users.

\subsection{Reinforcement learning}
\begin{figure} \centering
\includegraphics[width=0.3\textwidth]{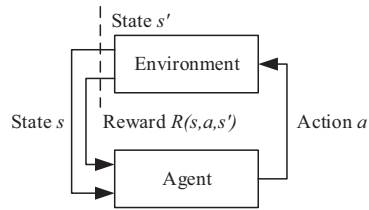}
\caption {Illustration of reinforcement learning. An agent interacts with its environment and receives the feedback in the form of rewards. The agent's objective is to learn to take actions to maximize the expected rewards.} \label{mdp}
\end{figure}

Reinforcement learning is very useful when little knowledge about an environment is known and a decision maker, i.e., an agent, needs to learn and adapt to its environment with significant uncertainty, such as the case of a wireless radio learning and adapting to the RF environment. As illustrated in Figure \ref{mdp}, a stochastic finite state machine is usually used to model the environment with inputs, such as an action sent from the agent, and outputs, such as observations and rewards sent to the agent. The agent's objective is to maximize rewards by exploring and exploiting the environment. Reinforcement learning can be applied to energy harvesting \cite{aprem2013transmit}, spectrum sensing \cite{reddy2008detecting}, and spectrum access in cognitive radio networks \cite{galindo2010distributed,venkatraman2010opportunistic}.

\emph{Markov decision process} (MDP) is a widely used mathematical framework to model decision-making under uncertainty, where the outcomes are partially random and partially controllable by an agent. 
To model a problem using MDP, the four components including the state space ($S$), the action space ($A$), the transition probabilities ($T$), and the reward functions ($R$), need to be specified. Several iterative algorithms, such as the value-iteration algorithm based on the Bellman's principle of optimality, can be used to identify the optimal action in each state. A \emph{partially observable Markov decision process} (POMDP) can be used to model the decision process in the situations where an agent does not directly observe the underlying states.

MDP is used to formulate the dynamic spectrum access problem \cite{berthold2008detection}, where the state space $S$ denotes all legitimate frequency bands that a secondary user can use. The set of available actions in each state $s \in S$ contains three types: performing a cycle of detection and transmission in the current frequency band $s \in S$, performing out-of-band detection in a different frequency band, and switching to another frequency band. Furthermore, a state transition occurs only if the action of switching to another band is selected. In addition, the reward function $R(s,a, s^{'})$ is defined as follows: the first type of rewards is the bits that have been transmitted while staying in the current frequency band $s \in S$, the second is the reward of performing a detection in a different frequency band, and the third is the reward of switching to another frequency band, which may be negative due to the transmission delay.

Going beyond traditional MDP, reinforcement learning does not require any prior knowledge of the transition probabilities $T$ or the reward functions $R$ and is capable of addressing complicated tasks when the traditional approaches are intractable.
This makes reinforcement learning a good fit for many real-world applications. Specifically, Q-learning is an efficient model-free reinforcement learning technique, which can identify an optimal policy for any finite MDP. 

Q-learning can be applied to spectrum access in cognitive radio networks. Specifically, Q-learning is used to control the interference in a cognitive radio network \cite{galindo2010distributed}, where the aggregated interference caused by the secondary users to the primary user is below a certain threshold. In particular, each secondary user needs to determine how much power it can transmit to avoid interference. The state of the secondary user consists of three components, a binary indicator specifying whether the secondary user interferes with the primary user, the estimated distance between the interference contour and the secondary user, and the power at which the secondary user is currently transmitting. The action set includes power levels of the secondary user and the reward due to action $a \in A$ in state $s \in S$ is quantified as the improvement in SINR. It is demonstrated that the strategy can alleviate the interference to the primary user regardless of the partial state observability.
In \cite{reddy2008detecting}, a reinforcement learning framework is proposed based on Q-learning to identify the presence of the licensed user signal and to access the licensed channels whenever they are idle. In \cite{xia2009reinforcement}, reinforcement learning based on Q-learning is used for routing in multi-hop cognitive radio networks, which allows learning the good routes efficiently.

A channel selection strategy based on multi-agent reinforcement learning is proposed in multi-user and multi-channel cognitive radio systems for secondary users to avoid the negotiation overhead \cite{li2009multi}. In contrast to single-agent reinforcement learning, there are more challenges associated with multi-agent reinforcement learning, such as the nonstationary and coordination.
In \cite{venkatraman2010opportunistic}, reinforcement learning is used to improve opportunistic spectrum access
in cognitive radio networks by interacting with the environment.

In \cite{jayaweera2011asymmetric}, reinforcement learning is employed in a dynamic spectrum leasing framework, which allows the proposed auction game to reach an equilibrium with both centralized and distributed network architectures.
A stochastic game framework based on reinforcement learning is proposed in \cite{wang2011anti} for anti-jamming defense. In particular, minimax Q-learning
is used to learn the optimal policy, which results in maximizing the expected sum of discounted payoffs defined as the spectrum-efficient throughput. Simulation results demonstrate that the optimal policy obtained from minimax Q-learning can achieve much higher throughput, in comparison with the myopic learning policy that maximizes the payoff at each step ignoring the dynamics of the environment.

A key component in reinforcement learning for cognitive radio networks is the tradeoff between exploration and exploitation. Specifically, novel exploration schemes, such as re-partitioning
and weight-driven exploration proposed in \cite{jiang2011efficient}, significantly outperform the traditional
uniform random exploration scheme.
A distributed multi-agent multi-band reinforcement learning framework is developed in \cite{lunden2011reinforcement} for spectrum sensing in ad hoc cognitive radio networks. The goal is to maximize the spectrum utilization for secondary use given a desired diversity order, where a desired number of secondary users can coexist in each frequency band. It is proved that the proposed model of spectrum sensing in a multi-agent scenario is computational efficient and can be deployed in networks with a large number of secondary users and a set of different frequency bands.
In \cite{bkassiny2011distributed}, a MAC protocol is proposed for autonomous cognitive radio users. The protocol is based on Q-learning and allows learning an efficient sensing policy in a multi-agent decentralized POMDP environment.

In contrast to full MDP where the environment has many states and new states depend on previous states and actions, \emph{multi-arm bandit} (MAB) can be regarded as a simple version of MDP where the environment has only one state. In this stateless situation, the reward depends only on the action, i.e., the arm, and the agent simply needs to learn to choose the best action, i.e., pull the arm, iteratively to maximize the sum of the cumulative rewards. MAB can be extended to a \emph{multi-player multi-armed bandit game} (MP-MAB), where the reward collected by any player depends on other players' decisions. MAB based models require the balance between choosing the actions to maximize rewards based on the acquired knowledge and attempting new actions to explore unknown knowledge, which is known as the aforementioned exploitation versus exploration tradeoff in reinforcement learning.

The MAB and MP-MAB models are capable of addressing channel selection problems in wireless communication systems, where some of the wireless environment parameters, such as the channel conditions, have to be ``explored" while the information of the known channels needs to be ``exploited". For example, a semi-dynamic parameter tuning scheme to update the multi-armed bandit parameters is proposed in \cite{alaya2008semi} to balance exploring the external environment and exploiting the acquired knowledge to decide which channel to access in dynamic environments.
The above machine learning models are summarized in Table~\ref{tab:sumMachine}.

\begin{table*}\centering
\caption{Comparison of Machine Learning Models for Intelligent Wireless Communications}\label{tab:sumMachine}
{
\centering
\begin{tabular}{|c|c|c|c|}
\hline
 Category & Learning Algorithms & Characteristics & Applications \\
\hline
\multirow{10}{*}{\parbox{1.9cm}{Supervised Learning}} & \multirow{2}{*}{KNN} & majority votes of neighbors & spectrum sensing \cite{thilina2013machine} \\
& & lazy learner & \\
\cline{2-4}
& \multirow{2}{*}{ Logistic/SVM-linear} & linear separable input & spectrum sensing \cite{thilina2013machine} \\
& & easy to train & MAC protocols \cite{yang2010mac,hu2014mac} \\
\cline{2-4}
& \multirow{2}{*}{SVM-nonlinear} & non-linear input to high dimension & spectrum sensing \cite{thilina2013machine} \\
& &expensive to train & MAC protocols \cite{yang2010mac,hu2014mac}\\
\cline{2-4}
& \multirow{2}{*}{Bayesian Net/HMM} & statistical models, interdependent outputs & spectrum sensing \cite{jiang2014multi,zheng2013spectrum}\\
& & such as Markov time series & channel estimation \cite{wen2015channel,choi2013estimation} \\
\cline{2-4}
& \multirow{2}{*}{ANN}& model any complicated function & spectrum sensing \cite{tang2010artificial,tumuluru2010neural,taj2011cognitive} \\
& & hard to train & signal classification \cite{popoola2011novel}\\
\hline \hline
\multirow{8}{*}{\parbox{1.9cm}{Unsupervised Learning}} & \multirow{2}{*}{k-means} & parametric, need to specify $k$ & network formation \cite{chen2007topology} \\
& &centroid based clustering, iteratively update & power optimization \cite{zhang2011distributed}\\
\cline{2-4}
& \multirow{2}{*}{DPMM }& nonparametric, clusters adapt to data & network clustering \\
& & fully Bayesian, approximate inference& \cite{shetty2009identifying,bkassiny2013multidimensional}\\
\cline{2-4}
& \multirow{2}{*}{PCA }& orthogonal axes to maximize variance & denoising \\
& & reduce dimension& \cite{hou2012spectrum,qiu2011cognitive} \\
\cline{2-4}
& \multirow{2}{*}{ICA} & independent components & source separation in smart grid \cite{qiu2011cognitive}\\
& & reduce dimension, signal separation &signal source decomposition \cite{nguyen2013binary}\\
\hline \hline
\multirow{7}{*}{\parbox{1.9cm}{Reinforcement Learning}} & \multirow{2}{*}{MDP/POMDP} & decision-making under uncertainty & energy harvesting \cite{aprem2013transmit} \\
& & specify full model ($S$, $A$, $T$, $R$) & dynamic spectrum access \cite{berthold2008detection}\\
\cline{2-4}
& \multirow{3}{*}{Q-learning} & unknown state transition and rewards & self-configuration in femtocells \cite{alnwaimi2015dynamic} \\
& & address complicated tasks efficiently & power control in small cells \cite{onireti2016cell}\\
& & & spectrum access \cite{onireti2016cell,reddy2008detecting,venkatraman2010opportunistic}\\
\cline{2-4}
& \multirow{2}{*}{Multi-arm bandit } & learning in stateless environment & channel selection\\
& & exploitation and exploration & \cite{alaya2008semi,maghsudi2015channel}\\
\hline
\end{tabular}
}
\end{table*}

\subsection{Emerging machine learning techniques}
Traditionally, the training process of a machine learning algorithm occurs in a centralized processor that contains all training examples. As more and more data are available, e.g., reaching petabyte or exabyte magnitude, distributed frameworks with parallel computing become a promising direction to scale up machine learning algorithms. Distributed computing platforms, such as Hadoop MapReduce and Spark, are developed to enable parallel computations on large clusters of machines. A general approach of implementing machine learning algorithms on top of MapReduce is investigated in \cite{chu2007map}. Another impressive progress is made in cloud-computing-assisted learning \cite{armbrust2010view}.

In cognitive and intelligent communications, most of the learning tasks, such as spectrum sensing, need to be finished within a certain period of time as observations change over time. In these time-sensitive cases, a learning algorithm needs to incorporate fresh input data and make predictions/decisions in a real-time manner. In contrast to the traditional offline or batch learning, which needs to collect the full training examples, the online learning \cite{shalev2012online}, a well-established learning paradigm, is capable of learning one instance at a time. In addition, several streaming processing architectures, including Borealis \cite{abadi2005design}, S4 \cite{neumeyer2010s4}, and Kafka \cite{goodhope2012building}, are proposed recently to support real-time data analytics \cite{yang2013big}.

Deep learning \cite{Lecun2015nature}, as one of the most popular research fields inspired by a large ANN, can capture complicated and potentially hierarchically organized statistical features of inputs and outperform state-of-the-art methods with carefully drafted hand-made features. Deep learning with either supervised or unsupervised strategies has shown great success in computer vision \cite{Krizhevsky2012imagenet}, speech recognition \cite{Hinton2012sig}, and natural language processing~\cite{Mikolov2013distributed}, as well as coding ~\cite{7886039}, signal detection and channel estimation ~\cite{DBLP:journals/corr/FarsadG17,8233654,8054694,ye2018power,8322184,8353153,wang2018deep,ye2018channel}, and resource allocation in wireless communications \cite{DBLP:journals/corr/SunCSHFS17}. Traditional neural networks with very few layers have been widely utilized in cognitive radio networks \cite{tsagkaris2008neural,tumuluru2010neural}. As the accelerated GPU computing becomes more and more popular, larger volumes of training data become available, and more and more effective training algorithms are developed, deep learning will play a pivotal role in supporting predictive analytics, which also makes it a promising research direction in supporting intelligent wireless communications.

\section{Applications in Wireless Systems, Challenges, and Future Directions} \label{app}
In this section, we focus on applications of cognitive radio and machine learning to the existing and future wireless communication systems. The compelling applications include small cells and heterogeneous networks, device-to-device communications, full-duplex communications, ultra-wideband millimeter-wave communications, and massive MIMO. For each of these applications, we discuss why intelligence is important, review how perception, reconfiguration, and machine learning techniques can be applied, and present technical challenges and future directions of cognitive radio technology and machine learning to further improve the level of intelligence and enable more applications.

\subsection{Small cells and heterogeneous networks}
The deployment of small cells, such as femtocells, has emerged as a promising technology to extend service coverage and increase network throughput \cite{chandrasekhar2008femtocell}. In a heterogeneous network, both small cells and macrocells face the cross-tier interference and co-tier interference from the network elements belonging to different and the same tiers, respectively. Intelligence is therefore important to improve the system performance for the coexistence of small cells and macrocells. First of all, the aforementioned spectrum sensing and environment perception techniques can be used in small cells to identify whether a macrocell is transmitting over a specific channel or not and facilitate interference management.

Meanwhile, spectrum-aware resource optimization can be designed in heterogeneous multicell networks \cite{salameh2016efficient}. In consideration of proportional fairness and traffic demands, the overall network spectrum efficiency can be maximized with the proposed channel allocation scheme. In \cite{wang2013resource}, resource allocation with imperfect spectrum sensing for heterogeneous OFDM-based networks is investigated. A two-step scheme decomposes the resource allocation problem into subchannel allocation and power allocation subproblems. While the optimal sum rate can be achieved through polynomial complexity, the near optimal sum rate can be achieved through constant complexity. From the energy efficiency perspective, centralized power allocation in heterogeneous networks is studied in \cite{ramamonjison2015energy}. The energy efficiency performance with exclusive spectrum use and spectrum sharing is optimized through Newton method based power allocation algorithms. The convergence of the proposed algorithms and the significant energy efficiency gains are illustrated. Applying graph-based and market-based approaches to the heterogeneous network users often involves an intermediary agent as a secondary service provider. In this way, small cells can purchase channels without spectrum sensing capabilities. This market structure is also designed for small cells with limited capability in computation and communications. In \cite{Li2016Auction}, an auction-based secondary spectrum market is proposed to share spectrum with small cell users in a two-tier heterogeneous network. In \cite{Cao2015Cognitive}, a virtual network operator is involved in a two-tier spectrum sharing market, in which users have heterogeneous demand requirements and channel valuations. By making the trading process a five-stage Stackelberg game, optimal decisions are proved to exist and an algorithm is proposed to make the optimal decisions. In \cite{Duan2013Economics}, the cellular operator provides both femtocell and macrocell services with limited spectrum resources. The spectrum allocation and pricing are modeled as a Stackelberg game and the optimal decisions under different assumptions are discussed.

Machine learning has been extensively applied to heterogeneous networks. For example, variations of SVM are adopted in \cite{yang2010mac} and \cite{hu2014mac} to classify MAC protocols, which is helpful for users to change their transmission parameters in heterogeneous networks. Meanwhile, small cells in heterogeneous networks can be clustered to avoid interference. A distributed strategy based on reinforcement learning is formulated as dynamic learning games in \cite{alnwaimi2015dynamic} for the optimization and self-configuration of femtocells in heterogeneous networks, where closed-access \emph{Long-Term Evolution} (LTE) femtocells overlay an LTE network. Specifically, the learning strategy enables opportunistic spectrum sensing of the radio environment and parameter tuning, to avoid the interference in heterogeneous networks and satisfy QoS requirements.
In dense small cell networks, Q-learning can be used to manage cell outage and compensation \cite{onireti2016cell}. The state space of the problem consists of the allocations of resources to users. The actions are downlink power control and the rewards are quantified as SINR improvement. It has been shown that the reinforcement learning based compensation strategy achieves better performance.

It is expected that more and more small cell stations will be deployed by end users in the future. As a result, it will be more difficult for the operators to manage them. Therefore, intelligent schemes with cognitive and learning capabilities need to be developed to allow better self-organization of user-deployed small cells.

\subsection{Device-to-device communications}
To further alleviate the huge infrastructure investment of operators, D2D communications has been considered as another promising technique for wireless communication systems \cite{doppler2009device,feng2013device,feng2014device}. Similar to the case of heterogeneous networks, intelligence is important for the coexistence of D2D and cellular links. The aforementioned spectrum sensing and environment perception techniques can also be used in D2D communications to facilitate interference management.

Furthermore, resource optimization can be used to achieve better performance in D2D communications. A message passing-based resource allocation scheme is designed in \cite{hasan2014distributed} to maximize the network throughput in a distributed manner. The spectrum efficiency of multi-user multi-relay networks is improved with the proposed resource allocation of low computational complexity. Meanwhile, the interference introduced by the coexistence of D2D and cellular users is effectively mitigated with the proposed interference and QoS constraints. In consideration of the channel uncertainties, the work in \cite{hasan2014distributed} is extended in \cite{hasan2015distributed}. A new distributed resource allocation scheme based on a stable matching approach is developed in the same framework. The achievable rate under channel uncertainties is improved in the D2D network. Since D2D communications introduce an alternative mode for users, resource allocation jointly optimized with mode selection is proposed to maximize the system throughput in \cite{yu2014joint}. Given different spectrum sharing patterns in different modes, the properties of different modes are explored through the proposed power control and channel assignment schemes. As a result, the spectrum efficiency is significantly improved. 

Machine learning has also been applied to distributed D2D communications \cite{maghsudi2015channel}, where each individual D2D user aims to optimize its own performance over the vacant cellular channels with unknown statistics to the user. The distributed channel selection problem is modeled as an MP-MAB game. Specifically, each D2D user is modeled as a player of the MP-MAB game while arms represent channels and pulling an arm corresponds to selecting a channel. A channel selection strategy, consisting of the calibrated forecasting and no-regret bandit learning strategies, is proposed. Recently, reinforcement learning is used for distributed resource allocation in D2D based vehicular networks \cite{DBLP:journals/corr/abs-1711-00968, DBLP:journals/corr/abs-1711-00968-2, 8345672,8345672-2}.

With the shift from capacity centric to content centric in wireless networks, D2D communications with caching can be a potential solution to content delivery. However, it is still difficult to enable low latency and provide better quality of experience. Therefore, intelligent schemes with cognitive and learning capabilities are promising for the self-organization of D2D communications by taking user behaviors into account.

\subsection{Full-duplex communications}
Most existing wireless systems use \emph{half-duplex} (HD) communications without exploring the potential of \emph{full-duplex} (FD) communications \cite{choi2010achieving}. With FD communications, a radio can transmit and receive signals over the same spectrum band simultaneously, which potentially doubles the spectrum efficiency. In comparison with traditional HD communications, the presence of self-interference in FD communications increases the design complexity. With intelligent spectrum sensing and resource optimization, the interference can be mitigated to approach the theoretical high spectrum efficiency. On the one hand, spectrum sensing and signal transmission of secondary users are proposed to occur at the same time for timely identification of channel occupancy status change and high-throughput transmission \cite{6913496}. On the other hand, matching theory is applied to the resource optimization and transceiver unit pairing in full-duplex networks \cite{6849223}.

To achieve simultaneous bidirectional communications, novel perception and reconfiguration techniques are needed to enable the coexistence of uplink and downlink while machine learning algorithms will be useful to further address the added complexity in interference management.

\subsection{Ultra-wideband millimeter-wave communications}
To keep up with the growing wireless traffic and applications, the future wireless communication systems will require not only higher spectrum efficiency but also more bandwidth resources. Wideband communications in the higher frequency band are therefore receiving more and more attention \cite{roh2014millimeter}. Intelligent spectrum sensing and resource optimization are important for such cases, especially in ultra-wideband millimeter-wave communications. While the aforementioned wideband spectrum sensing algorithms will naturally find ample applications, resource optimization is also important for the enhancement of spectrum and energy efficiency \cite{7961156}.

As the number of users and channels can be significantly higher than that in the existing communication systems, the scalability becomes extremely important. Besides the aforementioned wideband spectrum sensing algorithms, spectrum access and resource optimization need to be more efficient and adapt to the higher signal attenuation. For such a purpose, novel machine learning algorithms should be designed together with the use of perception capability and reconfigurability to intelligently achieve significantly better performance in future ultra-wideband millimeter-wave communication systems.

\subsection{Massive MIMO}
Massive MIMO uses a large number of antennas, usually at base stations, and can potentially bring huge improvements in throughput and energy efficiency \cite{lu2014overview,larsson2014massive}. To enable the exploitation of extra degrees of freedom provided by the excess of antennas, intelligence is especially important and can enhance both the performance and efficiency in the new scenario. 
A massive MIMO zero-forcing transceiver using time-shifted pilots is designed and analyzed in \cite{7008572}. Meanwhile, deep learning based channel estimation schemes for massive MIMO systems are proposed \cite{8322184,8353153}. To exploit the angular information, a spatial spectrum sharing strategy is proposed in \cite{7558167} for massive MIMO cognitive radio systems. Energy-efficient resource allocation is studied in \cite{7473828} for multi-pair massive MIMO systems.

As in the case of wideband communications, machine learning needs to be exploited to transform the big data from the extra degrees of freedom into the right data with improved perception capability and reconfigurability for future wireless communication systems with massive MIMO.

\section{Conclusions} \label{con}
The intelligence of cognitive radio and machine learning offers the potential to learn and adapt to the wireless environments. As the use of machine learning techniques in wireless communications is usually combined with cognitive radio technology, we have focused on both cognitive radio technology and machine learning to provide a comprehensive overview of their roles and relationship in achieving intelligent wireless communications. We have considered spectrum efficiency and energy efficiency, both of which are important characteristics of intelligent wireless communication systems. We have also presented some practical applications of these techniques to wireless communication systems, such as heterogeneous networks and D2D communications, and further elaborated some research challenges and likely improvements in future wireless communication systems.

\section*{Acknowledgement} \label{ack}
The authors would like to acknowledge the support from the National Science Foundation under Grants 1443894, 1560437, and 1731017, the Louisiana Board of Regents under Grant LEQSF (2017-20)-RD-A-29, and a research gift from Intel Corporation.

\bibliographystyle{IEEEtran}
\bibliography{IEEEabrv,ref}

\end{document}